\documentclass[a4paper]{article}
\usepackage{float}
\usepackage{graphicx} % Required for inserting images
\usepackage[margin=1in]{geometry}
\usepackage[colorlinks=true, allcolors=blue]{hyperref}

\usepackage[
    backend=biber,
    style=numeric-comp,
    sorting=none
]{biblatex}
\usepackage{amsmath,amssymb,amsfonts,braket}
\usepackage{xcolor}

\newcommand{\orcid}[1]{\href{https://orcid.org/#1}{\includegraphics[width=8pt]{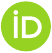}}}

\newcommand{\affil}[1]{{\fontsize{8}{10}\selectfont \raggedright #1 \\}}

\newcommand{\keywords}[1]{\vspace{3mm}{\fontsize{8}{10}\selectfont \raggedright {\bfseries Keywords:} #1}}

\newcommand{\email}[1]{\vspace*{12pt}{\fontsize{8}{10}\selectfont
       \raggedright {\bfseries E-mail:} \if@anonymous \phantom{#1} \else #1 \fi}
	  \vspace{3mm} }

\makeatletter
\renewcommand{\title}[1]{{\fontsize{18}{21}\selectfont\noindent\raggedright\textsf{#1}\par}\vspace{5mm}}
\renewcommand{\author}[1]{{\fontsize{10}{12}\selectfont\raggedright #1 \par}\vspace{3mm}}
\makeatother

\renewenvironment{abstract}{%
      \vspace{16pt plus3pt minus3pt}
      {\color{gray}\hrule} \ \\
      \noindent \fontsize{11}{12}\selectfont {\bfseries Abstract}\\
	  \rm\ignorespaces \raggedright}{\vspace{3mm} {\color{gray}\hrule}}

\begin{document}

%\maketitle

% 4. Your required metadata block
\title{Numerical Optimization of Two-Qubit Gates in Silicon Flip-Flop Qubit Arrays under Electrical Control}

\author{Lorenzo D'Onofrio$^1$\orcid{0009-0003-0317-6757}, Elena Ferraro$^1$\orcid{0000-0002-9844-2290} and Marco De Michielis$^{1,*}${\orcid{0000-0003-2988-3719}}}

\affil{$^1$CNR-IMM Unit of Agrate Brianza, Via Olivetti 2, Agrate Brianza, Italy}

\affil{$^*$Corresponding author: marco.demichielis@cnr.it}


\begin{abstract}
Silicon-based donor flip-flop qubits offer a promising path toward scalable, fault-tolerant quantum computing by combining the long coherence times of nuclear spins with fast, fully electrical control and long-range dipole-dipole coupling between qubits. However, realizing high-fidelity entangling operations in this platform remains challenging. The entangling interaction is intrinsically coupled to electron orbital dynamics, which can lead to leakage into non-computational states and unwanted phase accumulation. Furthermore, in multi-qubit architectures, residual dipolar couplings from spectator qubits distort the effective interaction landscape. In this work, we employ a numerical simulation framework, \texttt{FlipFlopQSim}, that models the spin-orbital dynamics of interacting flip-flop qubits to extract effective logical operations from realistic electrical control pulses. Using Makhlin invariants, we map the entangling landscape generated by electrically controlled dipole-dipole interactions and identify operating regions locally equivalent to canonical two-qubit gates, such as $\sqrt{\text{iSWAP}}$ and $\text{iSWAP}$. We then optimize the control parameters of physically realizable, electrically driven $R_z$ rotations to implement the necessary local corrections and maximize composite gate fidelity. Finally, we scale our analysis to multi-qubit registers with various geometries and connectivity patterns to evaluate spectator-induced distortions. Our results demonstrate that high-fidelity entangling operations cannot be optimized in isolation; rather, they require a co-design approach that simultaneously optimizes pulse control, local phase compensation, and physical device geometry. This work provides a robust numerical framework for assessing the scalability of electrically controlled silicon quantum processors and outlines key design principles for robust multi-qubit gate implementation.
% 240 words
\end{abstract}

\section{Introduction}

The realization of scalable quantum processors requires qubit platforms that combine long coherence times with fast, accurate, and technologically compatible control mechanisms. Silicon-based quantum devices represent a promising route toward fault-tolerant quantum computing due to their compatibility with mature semiconductor fabrication techniques ~\cite{Burkard-RevModPhys2023, DeMichielis-JPhysD-2023, McCallum-2021, Kane-1998, Vandersypen-2017, Morello-2020, Pla-2012, Pla-2013, Petta-2005, Laucht-2015, Muhonen-2014, Steger-2012, Tenberg2019SpinRelaxationMOS, Tyryshkin-2012}. Among the various silicon qubit architectures, donor-based flip-flop qubits provide a particularly attractive approach by encoding quantum information in the coupled electron-nuclear spin system of a $^{31}$P donor atom embedded in isotopically purified $^{28}$Si ~\cite{Tosi-2017, Tosi-2018, Savytskyy2023Science, Calderon-2022, Ferraro-2022, Simon-2020}.
This architecture combines the long coherence times of nuclear spins with the electrical controllability of electron wavefunctions, enabling fully electrical qubit manipulation without the need for oscillating magnetic fields, while also supporting long-range interactions via electric dipole-dipole coupling.

In flip-flop qubits, an applied electric field tunes the electron position between the donor site and a nearby Si/SiO$_2$ interface. This displacement modulates the hyperfine interaction, enabling electrically driven single-qubit operations through electric dipole spin resonance (EDSR). At the same time, the induced electric dipole moment provides a mechanism for long-range qubit coupling, forming the basis for electrically controlled two-qubit gates compatible with scalable semiconductor architectures.

However, realizing high-fidelity two-qubit operations in this platform presents several challenges. The entangling interaction is intrinsically linked to the electron orbital dynamics, requiring controlled shuttling of the electron wavefunction through different charge configurations. Consequently, gate performance depends not only on the strength of the dipolar coupling but also on electric transitions, leakage to non-computational states, and the accumulation of unwanted local phases. Furthermore, in realistic multi-qubit architectures, the presence of spectator qubits modifies the effective interaction landscape through residual dipolar couplings, making isolated two-qubit optimization insufficient for predicting scalable device performance.

Accurate evaluation of gate performance must account for the full physical Hilbert space containing both orbital and spin degrees of freedom, while simultaneously enabling simulations of increasingly large qubit systems. Moreover, the non-local nature of entangling operations must be separated from local single-qubit phases, since different physical implementations may belong to the same local-equivalence class while requiring distinct correction sequences to achieve a target computational gate.

In this work, we perform a numerical investigation of two-qubit gate optimization in silicon flip-flop qubits under low-complexity electrical control. We develop a simulation framework, \texttt{FlipFlopQSim}~\cite{FlipFlopQSim-2026}, capable of modeling the complete coupled spin-orbital dynamics of interacting flip-flop qubits and extracting the effective logical operations generated by realistic electrical pulse sequences, as outlined in Appendix~\ref{sec:FlipFlopQSim}. In Sec.~\ref{sec:FlipFlop}, we 
present the flip-flop qubit model, while in Sec.~\ref{sec:GateInfidelity}, we 
introduce the gate infidelity metric used to quantify the performance of the 
optimized operations. In Sec.~\ref{sec:Makhlin}, we characterize the entangling landscape generated by electrically controlled dipole-dipole interactions using the Makhlin invariants~\cite{Makhlin-2002} and identify operating regions locally equivalent to canonical two-qubit gates, including $\sqrt{\mathrm{iSWAP}}$ and $\mathrm{iSWAP}$ operations.
In Sec.~\ref{sec:Swaps}, we then investigate the implementation of the required local corrections through physically realizable electrically driven $R_z$ rotations, optimizing their control parameters to maximize the fidelity of composite two-qubit gates. 
Finally, in Sec.~\ref{sec:MultiQubit}, considering previous studies of multi-qubit silicon flip-flop qubit architectures with different geometries  \cite{DeMichielis-AQT-2024,DeMichielis_LA_SA_STA-AQT-2025}, we extend the analysis from isolated two-qubit systems to multi-qubit architectures with different geometries and connectivity patterns, investigating how the surrounding qubit environment modifies spectator-induced distortions. 

Our results show that high-fidelity entangling operations cannot be optimized solely by considering the intrinsic two-qubit interaction, but require a combined optimization of pulse control, local compensation, and device geometry.

These findings provide a numerical framework for assessing the scalability of electrically controlled flip-flop qubit architectures and highlight design principles for implementing robust two-qubit gates in silicon quantum processors.

\section{Flip-Flop Qubit Model}
\label{sec:FlipFlop}

In the flip-flop qubit architecture, quantum information is encoded in the joint spin state of the electron and the nucleus of a $^{31}\text{P}$ donor embedded in isotopically purified $^{28}\text{Si}$~\cite{Tosi-2017, Tosi-2018, Calderon-2022, Savytskyy2023Science, Ferraro-2022}. Both the electron and nuclear spins are spin-$1/2$ systems, characterized by gyromagnetic ratios $\gamma_e = 27.97\,\mathrm{GHz/T}$ and $\gamma_n = 17.23\,\mathrm{MHz/T}$, respectively. 

\begin{figure}[htbp!]
  \centering
  \includegraphics[width=0.6\linewidth]{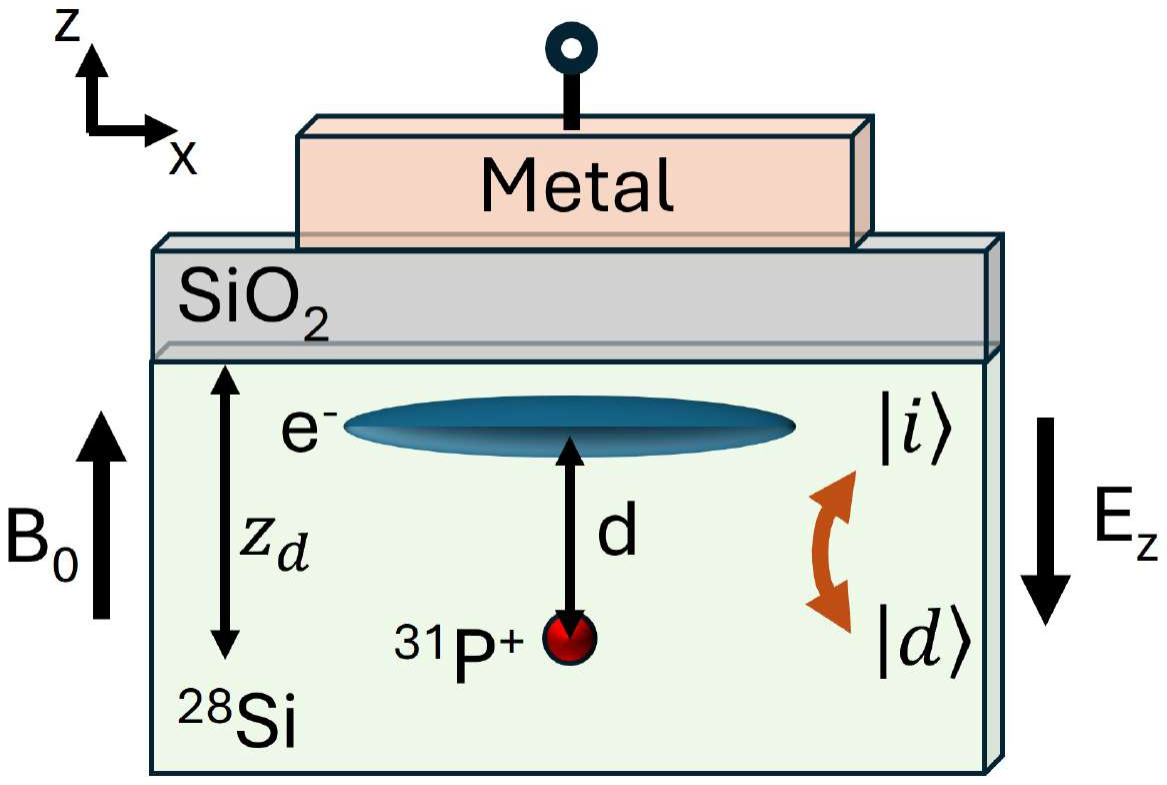}
  \caption{Schematic of a flip-flop qubit. A $^{31}\mathrm{P}$ donor atom is embedded in bulk $^{28}\mathrm{Si}$ at a distance $z_d$ from the Si/SiO$_2$ interface. An electric field $E_z$, applied through a metal gate, enables control of the electron position between the donor-bound state ($\ket{d}$) and the interface-bound state ($\ket{i}$). The parameter $d$ denotes the separation between the mean positions of the donor- and interface-bound electron wavefunctions. A static magnetic field $B_0$ is applied along the $\hat{z}$ axis.}
  \label{fig:DeviceSketch}
\end{figure}

The donor atom is positioned at a distance $z_d$ from a $\text{Si/SiO}_2$ interface. An applied vertical electric field $E_z$ controls the spatial position of the electron wavefunction between the donor nucleus and the interface, allowing dynamic tuning of both the hyperfine interaction $A(E_z)$ and the electron gyromagnetic ratio $\gamma_e(E_z)$. Specifically, $A(E_z)$ decreases from its bulk value $A_0 \approx 117\,\mathrm{MHz}$ to zero as the electron is pulled away from the donor toward the interface, while $\gamma_e$ can exhibit a relative variation of up to $\Delta\gamma \sim 0.007$~\cite{Tosi-2017}. A schematic of this device layout is shown in Fig.~\ref{fig:DeviceSketch}.

In the presence of a strong static magnetic field $B_0$ satisfying $(\gamma_e+\gamma_n)B_0 \gg A$, the system eigenstates are well approximated by product states of the electron and nuclear spins. The flip-flop transition frequency between the two antiparallel spin states defining the logical basis is then given by
\begin{equation}
\epsilon_{\mathrm{ff}} = \sqrt{\left[(\gamma_e(E_z)+\gamma_n)B_0\right]^2 + A(E_z)^2}.
\end{equation}
Modulating $E_z$ enables electrical control over $A(E_z)$, which drives transitions within the logical subspace via electric dipole spin resonance (EDSR).

To completely capture the system dynamics, we explicitly incorporate the orbital degree of freedom associated with the electron position ~\cite{Tosi-2017}. The total single-qubit Hamiltonian is expressed as
\begin{equation}
\label{eq:Hff}
\hat{H}_{\mathrm{ff}} = \hat{H}_{B_0} + \hat{H}_A + \hat{H}_{\mathrm{Orb}},
\end{equation}
where $\hat{H}_{B_0}$ represents the Zeeman term, $\hat{H}_A$ denotes the hyperfine interaction, and $\hat{H}_{\mathrm{Orb}}$ describes the tunneling and electric-field control within the eight-dimensional Hilbert space. Projecting this Hamiltonian onto the orbital eigenbasis $\{\ket{g},\ket{e}\}$, the logical qubit is encoded in the subspace $\{\ket{g\downarrow\Uparrow}, \ket{g\uparrow\Downarrow}\}$. 
Each term can be written in terms of the Pauli matrices
\begin{equation}
\hat{\sigma}_z = |g\rangle\langle g| - |e\rangle\langle e|,
\end{equation}
\begin{equation}
\hat{\sigma}_x = |g\rangle\langle e| + |e\rangle\langle g|,
\end{equation}
and the electron (nuclear) spin operators $\hat{\mathbf{S}}$ ($\hat{\mathbf{I}}$), with $z$-components $\hat{S}_z$ ($\hat{I}_z$).

The Zeeman Hamiltonian component reads
\begin{equation}
\hat{H}_{B_0} = \gamma_e B_0 \left[ \hat{\mathbb{I}} + \left( \frac{\hat{\mathbb{I}}}{2} + \frac{ d e \Delta E_z }{2 h \epsilon_o} \hat{\sigma}_z + \frac{V_t}{2 \epsilon_o} \hat{\sigma}_x \right) \Delta_{\gamma} \right] \hat{S}_z - \gamma_n B_0 \hat{I}_z,
\end{equation}
where $\Delta E_z = E_z - E_z^0$ represents the detuning of the vertical electric field from its value $E_z^0$ at the ionization point where the electron is shared equally between the donor and the interface. Here, $d$ is the separation between the mean positions of the donor-bound and interface-bound wavefunctions, $V_t$ is the tunnel coupling between the donor and interface wells, $e$ is the elementary charge, and $\epsilon_o$ denotes the orbital energy splitting
\begin{equation}
\epsilon_o = \sqrt{V_t^2 + \left(\frac{d e (E_z - E_z^0)}{h}\right)^2}.
\end{equation}

The hyperfine contribution is modulated by the electron configuration according to
\begin{equation}
\hat{H}_A = A \left( \frac{\hat{\mathbb{I}}}{2} - \frac{ d e \Delta E_z }{2 h \epsilon_o} \hat{\sigma}_z - \frac{V_t}{2 \epsilon_o} \hat{\sigma}_x \right) \hat{\mathbf{S}}\cdot\hat{\mathbf{I}},
\end{equation}
while the pure orbital and driving terms are gathered in
\begin{equation}
\hat{H}_{\mathrm{Orb}} = - \frac{\epsilon_o}{2}\hat{\sigma}_z - \frac{ d e E_a \cos(\omega_E t) }{2 h} \left( \frac{ d e \Delta E_z }{h \epsilon_o}\hat{\sigma}_z + \frac{V_t}{\epsilon_o}\hat{\sigma}_x \right),
\end{equation}
where $E_a$ and $\omega_E$ are the amplitude and angular frequency of the driving AC electric field, respectively. 

The dependence of the hyperfine coupling on the field detuning is captured by the parameterized profile
\begin{equation}
A(\Delta E_z) = \frac{A_0}{1 + \exp(c \cdot \Delta E_z)},
\end{equation}
where $c = 5.174 \times 10^{-4} \,\mathrm{m\,V^{-1}}$. Following the operational baseline established in Ref.~\cite{Tosi-2017}, we adopt the realistic system parameters $B_0 = 0.4\,\mathrm{T}$, $\Delta \gamma = -0.002$, and $d = 15\,\mathrm{nm}$ throughout this study. We also fix the tunnel coupling $V_t = 11.29\,\mathrm{GHz}$ in all gate simulations.

Multi-qubit coupling is mediated by a long-range electrostatic dipole-dipole interaction between the electric dipoles induced when the electrons are displaced toward the interface:
\begin{equation}
\label{eq:Hdip}
\hat{H}_{\mathrm{dip}}^{(i,j)} = \frac{1}{4\pi\varepsilon_0\varepsilon_{\mathrm{r}} r^3} \left[ \hat{\mathbf{p}}_i\cdot\hat{\mathbf{p}}_j - \frac{3(\hat{\mathbf{p}}_i\cdot\mathbf{r})(\hat{\mathbf{p}}_j\cdot\mathbf{r})}{r^2} \right],
\end{equation}
where $\varepsilon_0$ is the vacuum permittivity, $\varepsilon_{\mathrm{r}} = 11.7$ is the relative dielectric constant of silicon, and $\mathbf{r}$ is the displacement vector separating qubits $i$ and $j$. The effective dipole operator for the $k$-th qubit is given by
\begin{equation}
\label{eq:pk_0}
\hat{\mathbf{p}}_k = \frac{e d}{2} \left( \hat{\mathbb{I}}_k + \hat{\sigma}_{z,k}^{\mathrm{id}} \right),
\end{equation}
where the position operator takes the form
\begin{equation}
\label{eq:sigma_zID}
\hat{\sigma}_{z,k}^{\mathrm{id}} = \frac{d e \Delta E_{z,k}}{h\epsilon_o}\hat{\sigma}_{z,k} + \frac{V_t}{\epsilon_o}\hat{\sigma}_{x,k}.
\end{equation}
The total Hamiltonian governing an arbitrary network of $N$ interacting flip-flop qubits is thus obtained by summing the isolated single-qubit terms and all pairwise dipole-dipole contributions:
\begin{equation}
\hat{H} = \sum_{k=1}^{N} \hat{H}_{\mathrm{ff}}^{(k)}
+ \sum_{i=1}^{N-1}\sum_{j=i+1}^{N} \hat{H}_{\mathrm{dip}}^{(i,j)} .
\end{equation}

\section{Gate Infidelity}
\label{sec:GateInfidelity}

In this study, the performance of the resulting physical propagators is quantified using the gate infidelity $1-F$, restricted to the computational manifold~\cite{Pedersen-2007}:
\begin{equation}
\label{eq:GateFidelity}
F = \frac{1}{d(d+1)}\left[ \mathrm{Tr}(\mathcal{E} \mathcal{E}^\dagger) + \left|\mathrm{Tr}(\mathcal{E})\right|^2 \right],
\end{equation}
where $d = 2^N$ denotes the dimension of the $N$-qubit logical subspace, and $\mathcal{E}$ is the error operator projected onto it:
\begin{equation}
\mathcal{E} = P_{\mathrm{log}}\, U_{\mathrm{ideal}}^\dagger U_{\mathrm{ctrl}}\, P_{\mathrm{log}}.
\end{equation}
Here, $U_{\mathrm{ideal}}$ represents the target unitary gate, $U_{\mathrm{ctrl}}$ is the numerically simulated propagator driven by the time-dependent electric fields, and $P_{\mathrm{log}}$ is the projection operator onto the logical subspace encoding the quantum information.

The trace formulation in Eq.~\eqref{eq:GateFidelity} is mathematically equivalent to taking the uniform average of the state fidelity $|\langle \psi | U_{\mathrm{ideal}}^\dagger U_{\mathrm{ctrl}} | \psi \rangle|^2$ over the Haar measure for all pure input states $\ket{\psi}$ within the logical unit sphere $\mathcal{S}^{2d-1} \subset \mathbb{C}^d$~\cite{Pedersen-2007}. Restricting the error operator to the logical manifold ensures that our benchmarks evaluate only physically relevant computational states, thereby preventing virtual high-energy dynamics outside the computational subspace from introducing artificial fidelity penalties.

Evaluating gate performance across the unconstrained full Hilbert space would yield overly pessimistic metrics; for instance, phase accumulation on excited orbital states, while crucial as intermediate virtual pathways, can artificially degrade the global state overlap even when the net logical operation is flawless. In this restricted formulation, the first term in Eq.~\eqref{eq:GateFidelity} captures leakage out of the logical subspace, while the second term quantifies coherent deviations between $U_{\mathrm{ideal}}$ and $U_{\mathrm{ctrl}}$ within the logical manifold.

\section{Makhlin analysis and interaction landscape}
\label{sec:Makhlin}

Two-qubit entanglement is invariant under local transformations and is fully determined by local-unitary invariants. The nonlocal content of a two-qubit gate can therefore be characterized using the Makhlin invariants $(g_1, g_2)$ \cite{Makhlin-2002}, which uniquely classify equivalence classes of two-qubit unitaries up to local operations. These invariants provide a direct correspondence between an arbitrary two-qubit evolution and canonical entangling gates, including $\mathrm{CNOT}$, $\mathrm{iSWAP}$, and $\sqrt{\mathrm{iSWAP}}$.

In the flip-flop architecture, entanglement between qubits is mediated by dipole-dipole coupling. The interaction is activated dynamically through a controlled shuttling protocol in which both electron wavefunctions are adiabatically transferred from interface-localized states toward an ionization configuration where the dipolar coupling is enhanced. 

The protocol is implemented via a time-dependent detuning $\Delta E_z(t)$, which is swept from an idle value $\Delta E_{\mathrm{idle}}$, where the electron remains confined at the interface, to an intermediate value $\Delta E_{\mathrm{int}}$ over a time $\tau_1$. Subsequently, the system is driven over a time interval $\tau_2$ to the ionization point ($\Delta E_z = 0$~V/m), where the electron wavefunction is shared equally between the donor and the interface. The electrons are maintained in this configuration for a controlled interaction window $t_{\mathrm{hold}}$, after which the detuning is adiabatically ramped back to the initial operating point, restoring the original interface-localized states.

The resulting two-qubit unitary evolution depends parametrically on the hold time $t_{\mathrm{hold}}$ and the adiabatic ramp profiles defining the transfer sequence. Holding the ramp parameters fixed at $\Delta E_{\mathrm{idle}} = 10^4$ V/m, $\Delta E_{\mathrm{int}} = 1300$ V/m, $\tau_1 = 2 \, \mathrm{ns}$, and $\tau_2 = 20 \, \mathrm{ns}$, we analyze the non-local character of the gate as a function of $t_{\mathrm{hold}}$ for a two-qubit system with an inter-qubit distance of $r = 360\,\mathrm{nm}$. 

\begin{figure}[htbp!]
    \centering
    \includegraphics[width=0.90\textwidth]{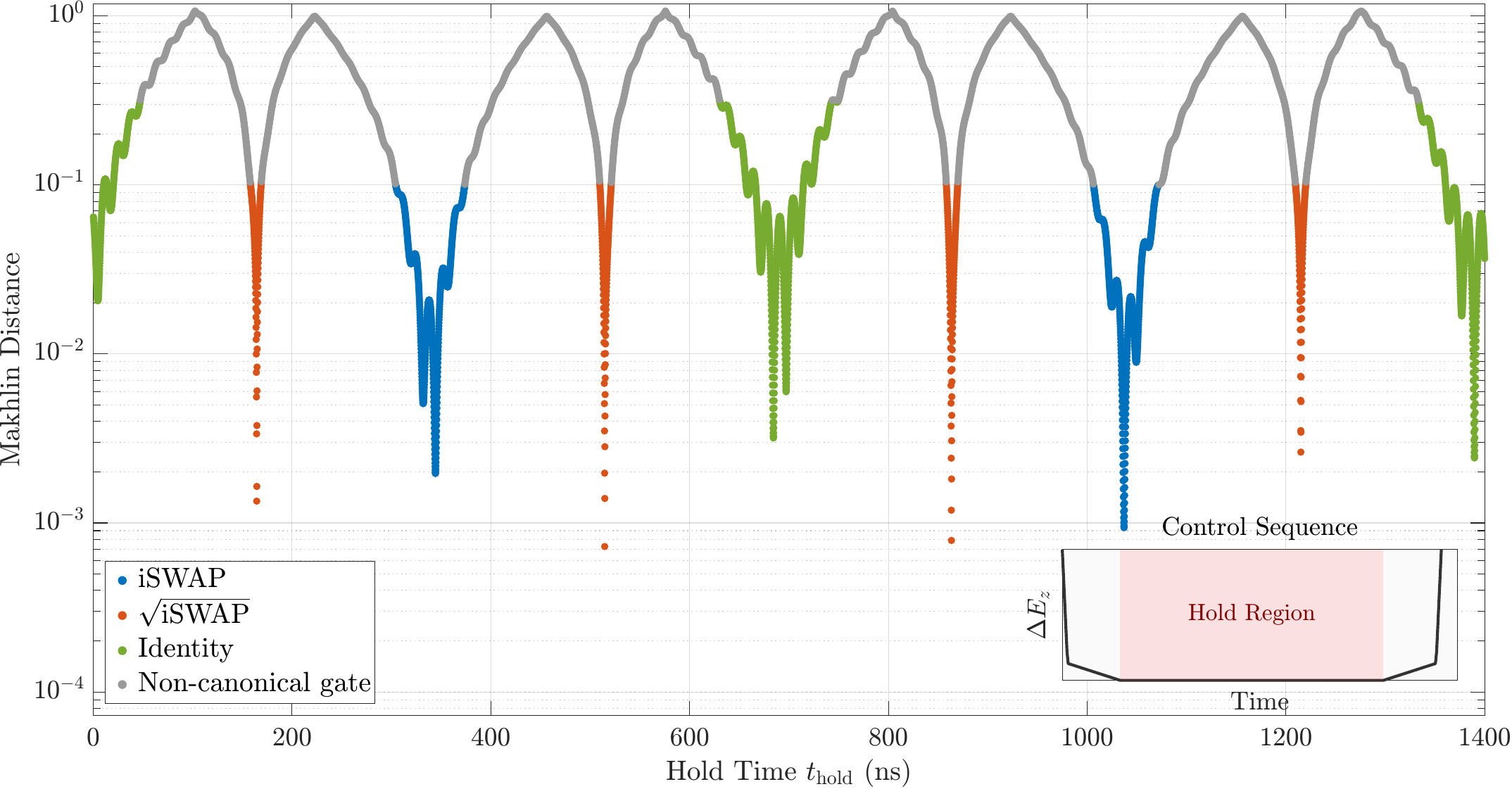}
    \caption{Classification of two-qubit operations via Makhlin invariants as a function of the hold time $t_{\mathrm{hold}}$ at the ionization point ($\Delta E_z = 0$~V/m). The inset shows the symmetric control sequence applied to both qubits, displaying the vertical electric field deviation $\Delta E_z$ from the ionization reference $E_z^0$. For each hold time, the plotted quantity is the minimum Euclidean distance between the computed Makhlin vector and the nearest ideal gate class ($\mathrm{iSWAP}$, $\sqrt{\mathrm{iSWAP}}$, $\mathrm{CNOT}$, or Identity). Data points are color-coded according to assigned gate class using the relative criterion $\|\mathbf{g}_\mathrm{ctrl}-\mathbf{g}_{\mathrm{ideal}}\| < 0.1\|\mathbf{g}_{\mathrm{ideal}}\|$.}
    \label{fig:makhlin_sweep}
\end{figure}

To classify the resulting gates, the Makhlin invariants are collected into a vector $\mathbf{g} = (g_1, g_2)$. As shown in Fig.~\ref{fig:makhlin_sweep}, we compute the minimum Euclidean distance between the extracted Makhlin vector $\mathbf{g}_{\mathrm{ctrl}}$ and the nearest canonical target gate class $\mathbf{g}_{\mathrm{ideal}}$ (specifically comparing against $\mathrm{iSWAP}$, $\sqrt{\mathrm{iSWAP}}$, $\mathrm{CNOT}$, and the Identity). A simulated gate is assigned to a local-equivalence class when this distance satisfies a relative threshold condition given by
\begin{equation}
\left\|\mathbf{g}_\mathrm{ctrl}-\mathbf{g}_{\mathrm{ideal}}\right\| < 0.1\,\left\|\mathbf{g}_{\mathrm{ideal}}\right\|.
\end{equation}
Applying this criterion reveals distinct intervals of $t_{\mathrm{hold}}$ where the evolution is locally equivalent to canonical $\sqrt{\mathrm{iSWAP}}$ and iSWAP gates.

The interaction landscape is periodic in the hold time $t_{\mathrm{hold}}$, with the underlying components of the extracted vector $\mathbf{g}_{\mathrm{ctrl}}$, specifically $\mathrm{Re}(g_1)$ and $g_2$, exhibiting periodicity of $\sim 720\,\mathrm{ns}$. Throughout the explored range of hold times, $\mathrm{Im}(g_1)$ remains close to zero, with absolute values below $10^{-3}$. Notably, no parameter region yields a CNOT-equivalent operation. This is consistent with the form of the dipole-dipole coupling Hamiltonian in Eq.~\eqref{eq:Hdip}, which reduces to an XY-type interaction in the orbital eigenbasis $\{\ket{g},\ket{e}\}$ and therefore cannot produce a CNOT gate within a single entangling step \cite{Schuch-2003}.

\section{Local Equivalence and Optimal Single-Qubit Corrections}
\label{sec:Swaps}

The resulting entangling unitaries can be mapped to $\sqrt{\mathrm{iSWAP}}$ and $\mathrm{iSWAP}$ gates via local $R_z$ rotations. Due to the periodic nature of the interaction landscape as a function of the hold time $t_{\mathrm{hold}}$, the same local equivalence classes are reached within multiple distinct time windows. 

In the two earliest windows in which the  $\sqrt{\mathrm{iSWAP}}$ and $\mathrm{iSWAP}$ equivalence classes are reached, the local correction takes the form of asymmetric pre- and post-$R_z$ operations:
\begin{equation}
\label{eq:UCorrected}
U_{\mathrm{corrected}} = \left( R_z(\theta) \otimes R_z(\pi + \theta) \right) \cdot U \cdot \left( I \otimes R_z(\pi) \right),
\end{equation}
where the rotation angles applied to the two qubits differ by $\pi$ and are parameterized by a single variable $\theta$. Consistent with the symmetry of the target gates, the assignment of these rotations to the two qubits can be interchanged, yielding an equivalent local correction:
\begin{equation}
\label{eq:UCorrectedVariant}
U_{\mathrm{corrected}} = \left( R_z(\pi + \theta) \otimes R_z(\theta) \right) \cdot U \cdot \left( R_z(\pi) \otimes I \right).
\end{equation}

The required local correction structure depends on the specific time window in which the equivalence class is reached. For example, in the second chronological window where a $\sqrt{\mathrm{iSWAP}}$ locally equivalent gate is obtained at a later hold-time interval, the correction simplifies to identical single-qubit rotations without a pre-rotation:
\begin{equation}
U_{\mathrm{corrected}} = \left( R_z(\theta) \otimes R_z(\theta) \right) \cdot U.
\end{equation}

In the following, we focus on the earliest time windows for both operations, as they provide the minimum interaction time at the ionization point while maintaining full gate reachability.

The optimal correction angle $\theta$ depends on the underlying interaction propagator and is thus a function of the hold time $t_{\mathrm{hold}}$.  Consequently, the local correction implementation must support tunable arbitrary $R_z$ rotations.

Within the flip-flop architecture, single-qubit $R_z$ operations are implemented using a control sequence analogous to that used for the entangling interaction. The electron is adiabatically transferred to the clock-transition bias point $\Delta E_{\mathrm{ct}} = 290$~V/m, where dephasing is minimized. A double-ramp shuttling sequence is applied, followed by a variable hold time $t_{\mathrm{hold}}$, after which the system is returned to the interface configuration. The dynamical phase accumulated during this hold interval directly defines the $Z$-rotation angle.

\begin{figure}[htbp!]
    \centering
  \includegraphics[width=0.90\textwidth]{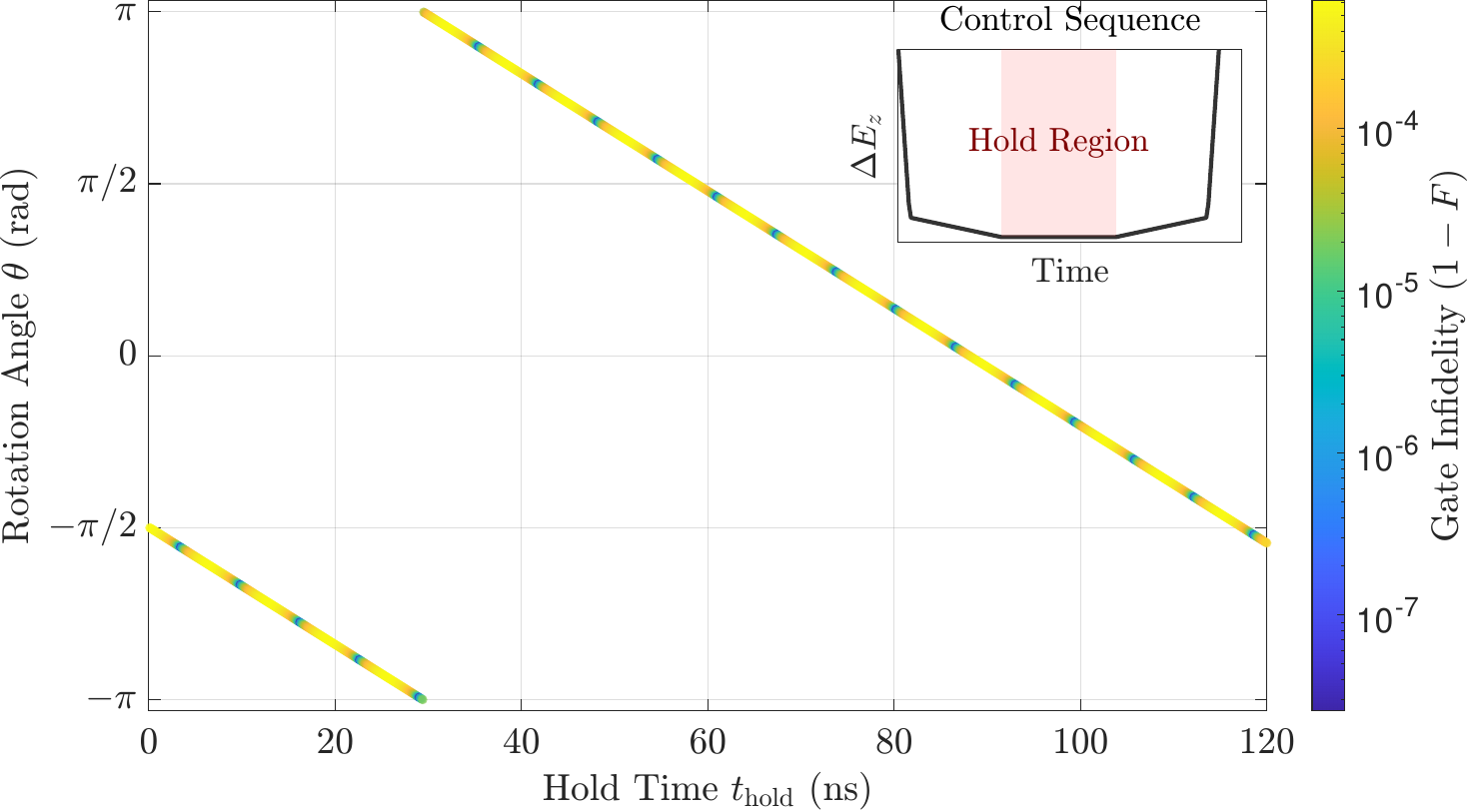}
    \caption{Phase calibration landscape for an $R_z(\theta)$ gate in a single-qubit system. 
    The rotation angle $\theta$ (y-axis) is extracted via bounded optimization of the effective 
    propagator as a function of the hold time $t_{\mathrm{hold}}$ at the clock-transition point (x-axis). 
    Color mapping indicates gate infidelity ($1-F$), accounting for both coherent errors and leakage to higher energy levels. The inset shows the control sequence, displaying the vertical electric field deviation $\Delta E_z$ from the ionization reference $E_z^0$. Ramps are executed to and from the clock-transition bias point $\Delta E_{\mathrm{ct}} = 290\,\mathrm{V/m}$.}
    \label{fig:rz_calibration}
\end{figure}

As shown in Fig.~\ref{fig:rz_calibration}, we find that a hold-time interval of $t_{\mathrm{hold}} \in [0, \sim 118\,\mathrm{ns}]$ spans the full $2\pi$ rotation phase. The calibration landscape exhibits high fidelity, with all operating points remaining below the fault-tolerance threshold of $1-F \leq 10^{-3}$ and selected configurations reaching infidelities as low as $10^{-7}$ and below. The target rotation angle $\theta$ exhibits a linear dependence on the hold time, resulting from the constant difference between the qubit precession frequency at the clock-transition bias point $\Delta E_{\mathrm{ct}} = 290 \,\mathrm{V/m}$ and at the idle configuration. Superimposed on this linear phase-accumulation behavior is a periodic fine structure in the infidelity landscape, which depends non-trivially on both $t_{\mathrm{hold}}$ and the specific ramp parameters due to transient effects during the shuttling sequence.

Modifying the ramp profiles reshapes this fine structure while preserving full $2\pi$ phase coverage within the same operational window. This structural flexibility allows us to optimize gate fidelity for any target rotation angle by jointly tuning the ramp parameters and the hold time.

To find these optimal configurations, we map the parameter space by sweeping $(\tau_1, \tau_2)$ over a uniform grid with a resolution of $0.1\,\mathrm{ns}$ across the ranges $[0.1, 5.0]\,\text{ns}$ and $[10, 20]\,\text{ns}$, respectively, while keeping $\Delta E_{\mathrm{idle}} = 10^4\,\mathrm{V/m}$ and $\Delta E_{\mathrm{int}} = 1300\,\mathrm{V/m}$ fixed. For each pair, the full dependence on $t_{\mathrm{hold}}$ is sampled. This dataset yields the optimal parameter triple $(\tau_1, \tau_2, t_{\mathrm{hold}})$ for any target rotation angle $\theta$.
%a procedure applied to both isolated single-qubit operations and two-qubit configurations with an idle spectator qubit.

\begin{figure}[htbp!]
    \centering
\includegraphics[width=0.90\textwidth]{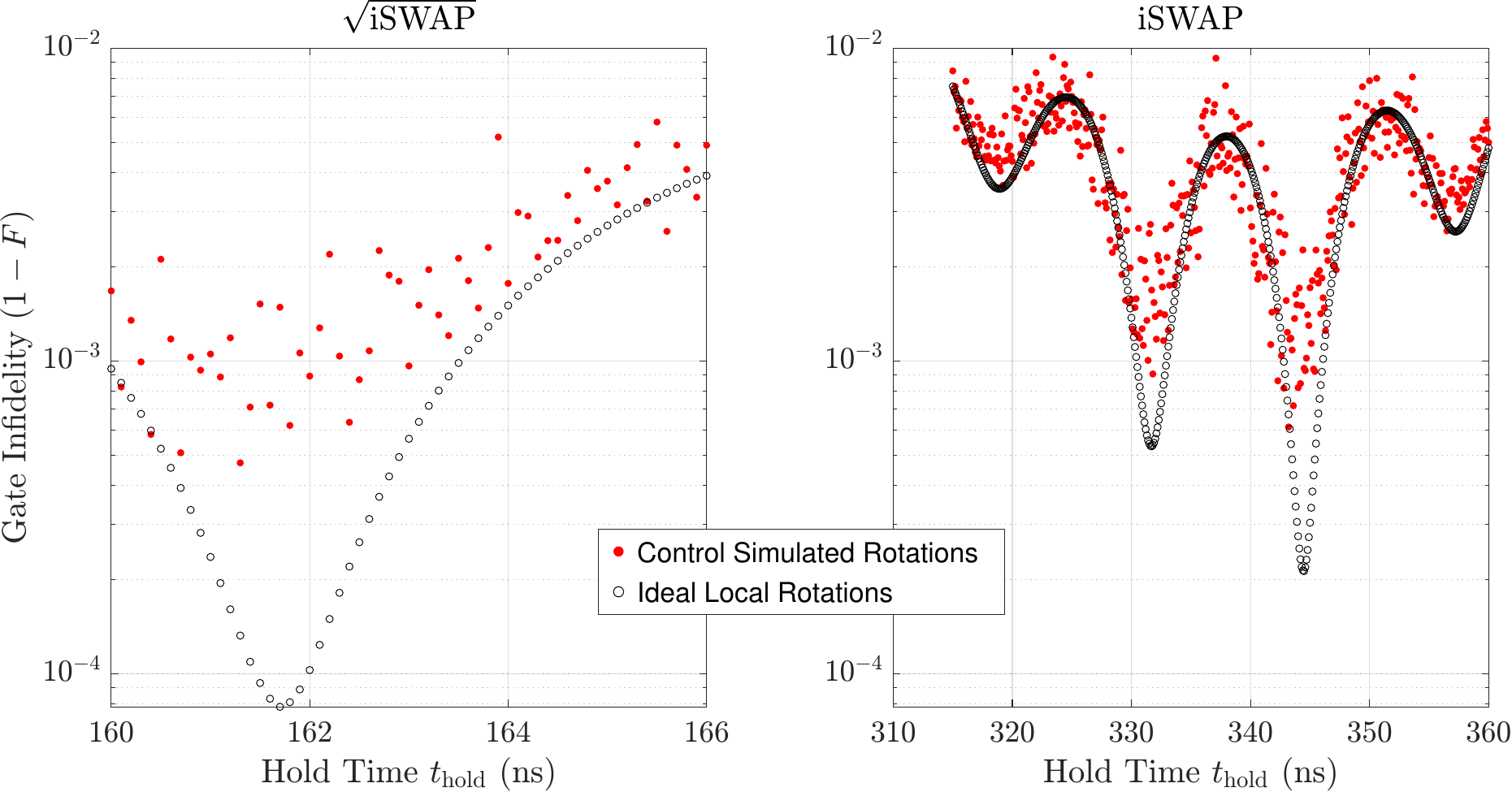}
 \caption{Optimization landscape for composite two-qubit entangling gates. Gate infidelity is shown as a function of the hold time $t_{\mathrm{hold}}$ at the ionization point ($\Delta E_z = 0$~V/m) for $\sqrt{\mathrm{iSWAP}}$ and iSWAP targets. Hollow points show the infidelity obtained using ideal, instantaneous $R_z$ corrections, while red points show the corresponding results using physically simulated electrically driven $R_z$ operations extracted from the calibration dataset. Both implementations reach the fault-tolerance threshold at their optimal operating points.}
    \label{fig:swap_grid_search}
\end{figure}

Using these optimal configurations, optimized pulse sequences implementing iSWAP and $ \sqrt{\mathrm{iSWAP}} $ gates are obtained.
In particular, for $ t_{\mathrm{hold}} $ regions where the Makhlin analysis indicates local equivalence to $ \sqrt{\mathrm{iSWAP}} $ and iSWAP, the optimal correction angle $ \theta $ is determined by minimizing the infidelity of the corrected gate under idealized, instantaneous, and error-free $ R_z $ rotations. The corresponding control parameters for the numerically simulated electrically driven $R_z$ operations are then retrieved from the calibrated $R_z$ dataset, and the full physically simulated gate sequence is reconstructed to evaluate the resulting infidelity. The $R_z$ operations in Eq.~\eqref{eq:UCorrected} are applied sequentially, with only one qubit driven at a time while the other remains idle. The results are shown in Fig.~\ref{fig:swap_grid_search}.

The physical corrected landscape follows the structure of the ideal corrected landscape, but the minimum infidelities are not reproduced exactly due to imperfections in the implemented $R_z$ rotations.
The optimal operating points of both composite gate implementations reach the fault-tolerance threshold, with infidelities below $1-F \leq 10^{-3}$ for both ideal and physically simulated $R_z$ corrections. In particular, the full 
physical simulations achieve infidelities of $4.72 \times 10^{-4}$ for the 
$\sqrt{\mathrm{iSWAP}}$ gate and $6.14 \times 10^{-4}$ for the 
$\mathrm{iSWAP}$ gate. The calibrated electrically driven single-qubit rotations therefore provide an accurate compensation of the local phases accumulated during the entangling evolution.

%The corrected physical gate exhibits strong sensitivity to the correction angle, with deviations as small as $10^{-4}$ radians leading to a degradation in fidelity. Consequently, precise tuning of $ \theta $ is essential. Operating even slightly away from the optimal correction angle consistently increases the gate infidelity, even when the underlying $R_z$ implementation has higher fidelity. This shows that improving the individual single-qubit rotations alone cannot compensate for an inaccurate local correction angle, indicating that the optimum is sharply defined.

The corrected physical gate exhibits strong sensitivity to the correction angle, with deviations as small as $10^{-4}$ rad leading to a degradation in fidelity. Consequently, precise tuning of $\theta$ is essential. When operating away from the optimal correction angle, the physical corrected-gate landscape deviates from the ideal baseline, resulting in increased infidelity across the vast majority of configurations, even for small angular offsets and when the underlying $R_z$ implementation exceeds the fidelity achieved at the optimal correction angle. This highlights that accurate calibration of the local phase correction is critical for recovering high-fidelity two qubit gates.

\section{Extension to Multi-Qubit Arrays}
\label{sec:MultiQubit}
 
When extending the optimized $\sqrt{\mathrm{iSWAP}}$ and iSWAP sequences to multi-qubit architectures, where a single pair of nearest-neighbor qubits is actively driven while all surrounding spectator qubits remain idle, the isolated two-qubit framework becomes insufficient. The main limitation arises from the modified effective entangling interaction landscape in the presence of spectator qubits. This modification alters both the nonlocal evolution and the local phases accumulated during the entangling process, shifting the local-equivalence regions away from those identified in the isolated two-qubit system. Since the optimal $R_z$ correction angle $\theta$ is directly linked to the local phases accumulated during the entangling evolution, the value calibrated in the two-qubit setting is no longer optimal in larger architectures, leading to a mismatch between the entangling dynamics and the applied local phase compensation. The calibrated $R_z$ operations themselves retain high fidelity for a given target angle; however, the required correction angle must be recalibrated to account for the modified evolution. Furthermore, the presence of spectator qubits generally results in an increased minimum achievable distance from the ideal Makhlin invariants for both $\sqrt{\mathrm{iSWAP}}$ and iSWAP, with a more pronounced effect for iSWAP. Therefore, two-qubit gate optimization in multi-qubit systems requires accounting for the environment-dependent entangling dynamics rather than relying on isolated two-qubit calibrations.

To construct the effective two-qubit description within the multi-qubit environment, we extract the unitary operator acting on the active pair by projecting the idle spectator qubits onto their $|0\rangle$ ground states. The Makhlin invariants are subsequently evaluated from this reduced operator. This projection implicitly restricts our analysis to conditional trajectories where the spectator qubits remain unexcited at the final gate time. This approach is required to preserve the unitariness of the reduced operator, as a partial trace over the idle degrees of freedom would generally introduce non-unitarity and mixed-state dynamics for which the invariants are undefined. 

Because the $t_\mathrm{hold}$ regions corresponding to $\mathrm{iSWAP}$ gates exhibit extreme sensitivity to variations in geometry and system size, we focus our multi-qubit analysis exclusively on the more resilient $\sqrt{\mathrm{iSWAP}}$ gate and consider only the earliest time window where local equivalence is achieved.

%\textcolor{violet}{To evaluate how these environmental deformations manifest in the final gate performance, we analyze the composite gate fidelity after incorporating the local phase adjustments. Embedding the entangling interaction within a composite sequence of optimized local $R_z$ corrections reveals that the resulting infidelity profile as a function of $t_{\mathrm{hold}}$ is significantly less steep than the underlying Makhlin-distance landscape. Low-infidelity regions extend beyond the narrow boundaries of strict mathematical local equivalence; while both metrics minimize in nearby parameter regions, their absolute extrema do not generally coincide. A prominent exception to this trend occurs in a five-qubit surface-code-inspired geometry, consisting of four edge qubits arranged in a square around a central qubit, during an interaction between the central qubit and an edge qubit. Here, the minimum of the corrected-gate infidelity shifts substantially away from the regions prescribed by the Makhlin invariants. This discrepancy underscores that local-equivalence analysis alone does not fully capture achievable gate performance. Even when the non-local content of the raw unitary diverges from the ideal target, joint optimization of the local $R_z$ corrections at a fixed $t_{\mathrm{hold}}$ can compensate for the environment-induced distortion to yield high-fidelity operations.}

To evaluate how the presence of spectator qubits affect the final gate performance, we analyze the composite gate fidelity after incorporating the optimized local phase adjustments and determine the minimum achievable infidelity across different qubit layouts.
We classify the layouts according to the neighborhood topology of the active qubit pair into \emph{symmetric} and \emph{asymmetric} configurations. In symmetric layouts, both active qubits experience equivalent local environments, with matching numbers of neighbors and identical coupling networks, whereas asymmetric layouts introduce an imbalance in the crosstalk and interaction landscape experienced by the active qubits.

For each device geometry and layout, in the $t_{\mathrm{hold}}$ interval where the Makhlin analysis identifies local equivalence to $\sqrt{\mathrm{iSWAP}}$, the correction angle $\theta$ is optimized by minimizing the infidelity of the corrected gate (Eq.~\ref{eq:UCorrected}) assuming ideal, instantaneous, and error-free $R_z$ rotations. From the resulting corrected gates, we select the optimal one and its corresponding correction angle $\theta$. The control parameters corresponding to this optimized $\theta$ are then retrieved from the calibrated dataset for the numerically simulated electrically driven $R_z$ operations, and the complete pulse sequence is reconstructed to evaluate the resulting infidelity. Note that the infidelity obtained with the physically simulated $R_z$ operations in this procedure does not, in general, coincide with the minimum achievable infidelity obtained from a joint optimization of the entangling interaction hold time and the correction angle $\theta$ under realistic pulse implementations.

\begin{figure}[htbp!]
    \centering
    \includegraphics[width=0.90\textwidth]{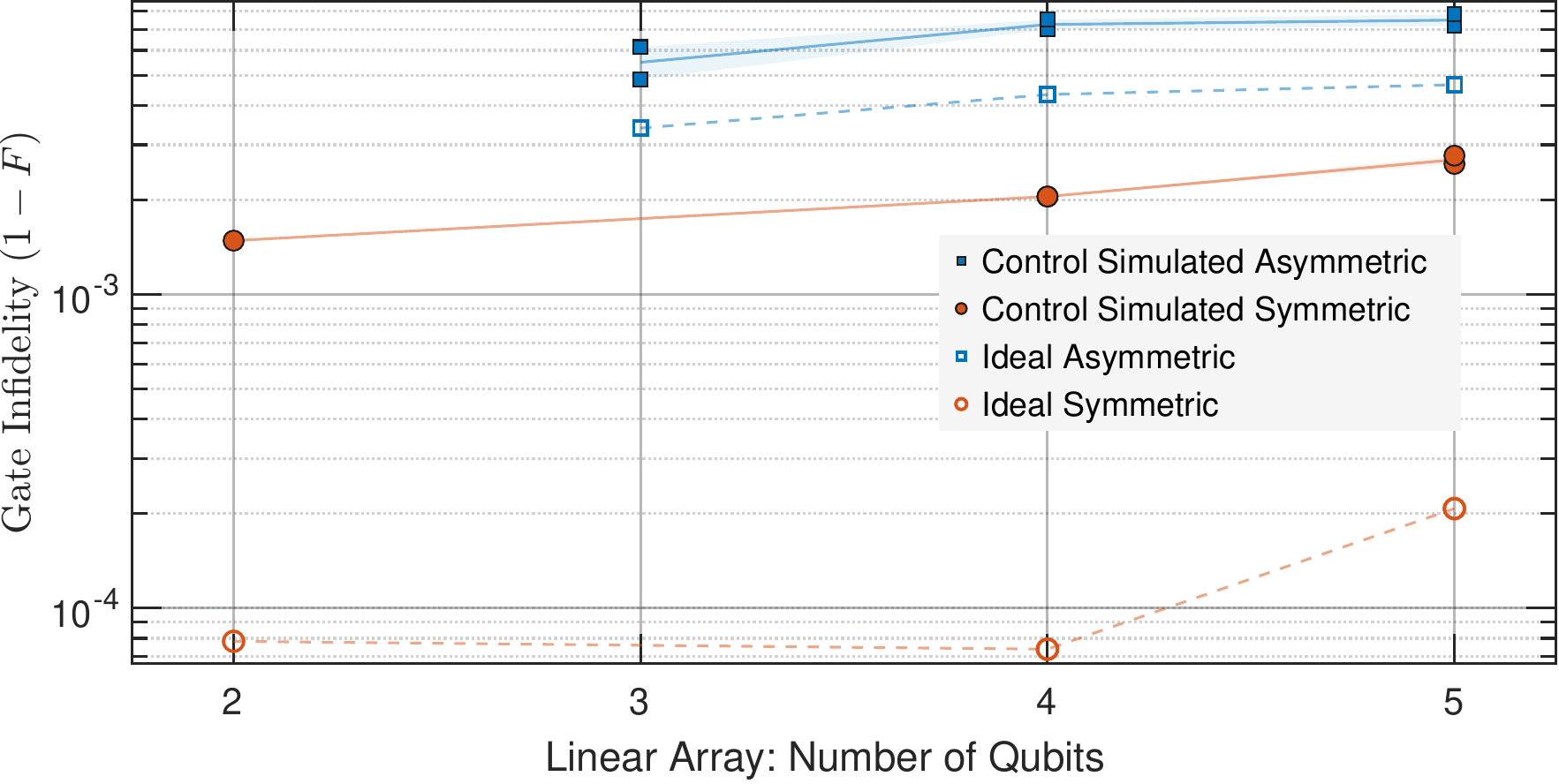}
        \caption{Lowest gate infidelity for $\sqrt{\mathrm{iSWAP}}$ gate in linear arrays with varying qubit numbers and active-idle layouts. Symmetric and asymmetric configurations are compared using both ideal $R_z$ corrections (instantaneous and error-free) and numerically simulated electrically driven $R_z$ gates obtained from the calibrated parameter dataset. Symmetric configurations consistently achieve lower infidelity than asymmetric layouts across the considered systems.}
    \label{fig:linear_array_infidelity}
\end{figure}

As shown in Fig.~\ref{fig:linear_array_infidelity}, we investigate linear arrays with a fixed inter-qubit distance $r = 360 \,\mathrm{nm}$, varying the number of qubits and the active-idle configurations. Across all system sizes, symmetric configurations consistently achieve lower minimum infidelities than asymmetric ones, indicating that asymmetric environments amplify the detrimental effects of spectator-qubit interactions. Notably, symmetric layouts with realistic calibrated physical $R_z$ pulses outperform asymmetric configurations even when the latter are evaluated assuming ideal, instantaneous corrections. This shows that asymmetric layouts substantially degrade gate performance, emphasizing the role of the surrounding interaction network in multi-qubit gate optimization.

\begin{figure}[htbp!]
    \centering
    \includegraphics[width=0.90\textwidth]{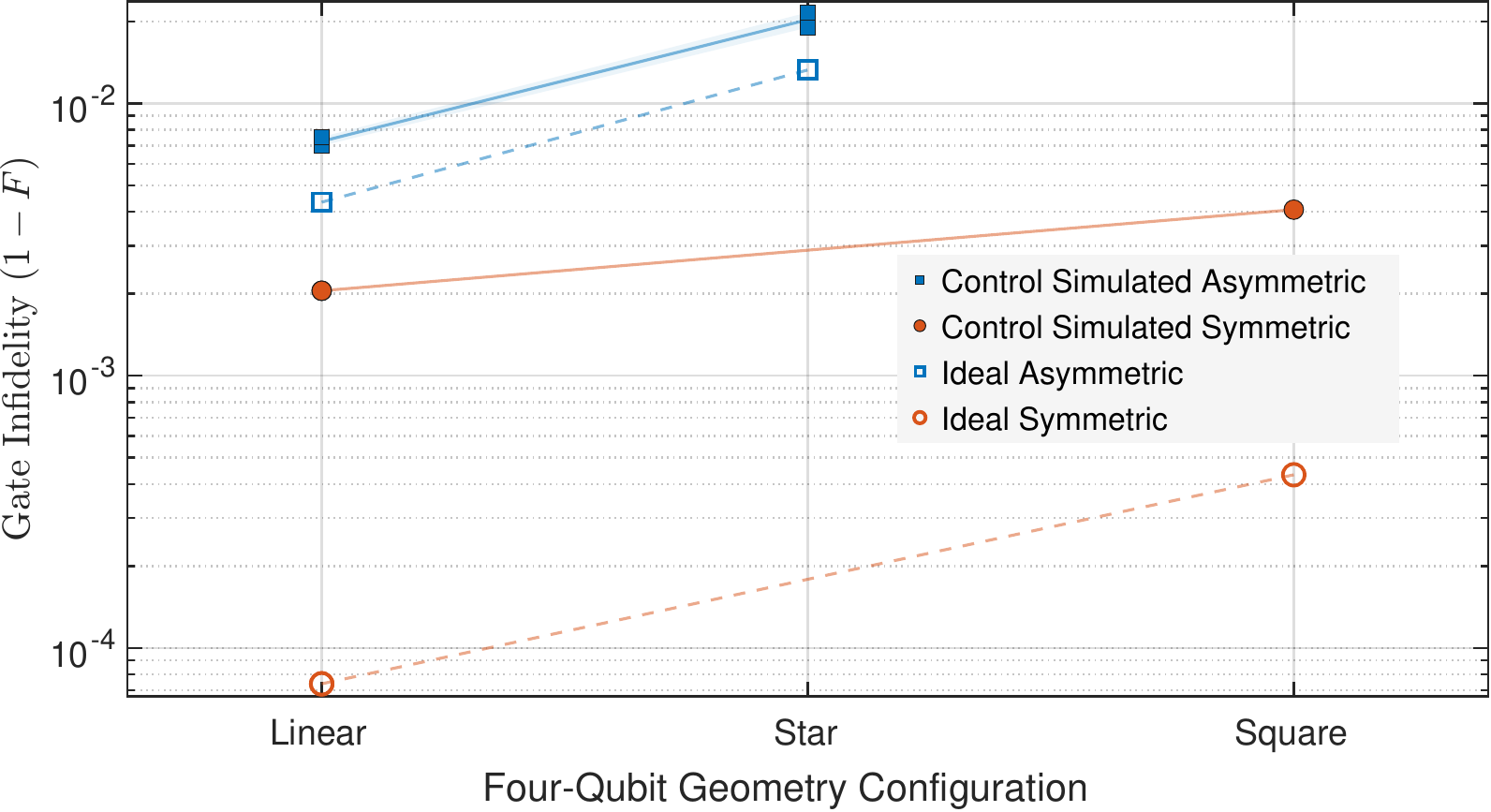}
    \caption{Lowest gate infidelity for $\sqrt{\mathrm{iSWAP}}$ gate across different four-qubit geometries and active-idle configurations. Symmetric and asymmetric layouts are compared using both ideal $R_z$ corrections (instantaneous and error-free) and realistic electrically driven $R_z$ implementations obtained from the calibrated parameter dataset. Symmetric configurations consistently achieve lower infidelity across the considered geometries.}
    \label{fig:4qubitGeometryBestInfidReal}
\end{figure}

This behavior extends beyond one-dimensional geometries. Figure~\ref{fig:4qubitGeometryBestInfidReal} compares different four-qubit layouts: linear, square, and star configurations. The linear geometry consists of equally spaced qubits with a inter-qubit distance $r = 360 \,\mathrm{nm}$. The square geometry consists of four qubits placed at the vertices of a square of side length $r = 360\,\mathrm{nm}$, with the active qubits occupying adjacent vertices. The star geometry consists of a central qubit coupled to three neighboring spectator qubits arranged at $120^\circ$ intervals, with a nearest-neighbor distance $r = 360\,\mathrm{nm}$. Across these architectures, symmetric connectivity patterns continue to exhibit lower gate infidelity.

In these simulations, the $R_z$ corrections can be implemented using either of the two equivalent local-correction forms reported in Eqs.~\eqref{eq:UCorrected} and \eqref{eq:UCorrectedVariant}. Consequently, the controlled gate simulation may yield two solutions for the same operating point, corresponding to the two possible assignments of the single-qubit rotations.
These solutions are equivalent under exchange of the qubit labels, and their negligible difference confirms that the implementation fidelity of the local $R_z$ corrections is largely insensitive to the device geometry.
%This contrasts with the entangling evolution, whose nonlocal dynamics are strongly modified by the surrounding qubit architecture.

\clearpage

\section{Conclusions and perspectives}
\label{sec:Conclusions}

% ===== sqrt(iSwap)  Effective Gate Infidelity: 4.716014e-04 ===== 
% ===== iSwap        Effective Gate Infidelity: 6.141524e-04 ===== 
 
In this work, we have performed a detailed numerical investigation of two-qubit gate optimization in silicon flip-flop qubits under low-complexity electrical control. Using a dedicated numerical simulation framework, \texttt{FlipFlopQSim}~\cite{FlipFlopQSim-2026}, capable of capturing the coupled spin-orbital dynamics, we accurately extracted effective logical gate operations generated by realistic, time-dependent electrical control pulses. Through the application of Makhlin invariants, we mapped the entangling landscape generated by electrically controlled dipole-dipole interactions, identifying parameter regions locally equivalent to canonical $\sqrt{\text{iSWAP}}$ and $\text{iSWAP}$ gates. We showed that while the intrinsic interaction generates entangling gates, achieving high-fidelity operations requires active compensation of local phase accumulations. By optimizing the control parameters of physically realizable, electrically driven $R_z$ rotations, we showed that these unwanted local phases can be efficiently corrected to yield composite gate infidelities below the fault-tolerance threshold of $1-F \leq 10^{-3}$. In particular, the optimum $\sqrt{\mathrm{iSWAP}}$ and $\mathrm{iSWAP}$ gates 
obtained from the full physical simulation achieve infidelities of 
$4.72 \times 10^{-4}$ and $6.14 \times 10^{-4}$, respectively. Furthermore, we extended our analysis to multi-qubit architectures to evaluate how idle spectator qubits modify the effective interaction landscape. Our results reveal that residual dipolar couplings from neighboring qubits significantly alter both the nonlocal evolution and local phase accumulation. Consequently, two-qubit gate optimization cannot be treated as an isolated two-qubit problem. Instead, it demands a coordinated co-design approach that simultaneously optimizes pulse control, local phase compensation, and physical device geometry. In particular, our comparison of different multi-qubit layouts, including linear, square, and star configurations, shows that symmetric qubit environments are highly beneficial, consistently yielding lower gate infidelities than asymmetric configurations under realistic control constraints.

Looking forward, an important research direction is the development of strategies to actively control spectator-induced couplings. In particular, future work will investigate the integration of dynamically tunable metallic couplers capable of effectively switching inter-qubit interactions on and off. Beyond interaction control, such couplers can extend the effective range of dipole–dipole coupling, allowing for increased spatial separation between donor qubits. This enlarged footprint relaxes geometric constraints on device layout, facilitating the routing of control lines and the integration of classical electronics without sacrificing qubit connectivity. The combination of these architectural solutions with the co-optimization of pulse sequences and local corrections identified in this work provides a concrete pathway toward scalable, high-fidelity silicon quantum processors.

\paragraph{Code Availability}
\label{sec:CodeAvailability}
\texttt{FlipFlopQSim}~\cite{FlipFlopQSim-2026}, the in-house MATLAB simulation software developed for this study, is openly available in the Zenodo repository at \href{https://doi.org/10.5281/zenodo.21669875}{10.5281/zenodo.21669875}.

\paragraph{Data Availability}
\label{sec:DataAvailability}
The data that support the findings of this study are available from the corresponding author upon reasonable request.

\appendix
\section{FlipFlopQSim}
\label{sec:FlipFlopQSim}

%To simulate the dynamics of the multi-qubit flip-flop architecture under time-dependent electric fields, we developed a numerical framework optimized to handle the exponential growth of the multi-body Hilbert space.

To simulate the coherent dynamics of multi-qubit flip-flop architectures under time-dependent electric control, we employ an in-house MATLAB framework, \texttt{FlipFlopQSim}~\cite{FlipFlopQSim-2026} (see \nameref{sec:CodeAvailability} for availability details). The framework provides a modular workflow encompassing device geometry and physical parameter definitions, Hamiltonian construction, control pulse generation and routing, dynamical propagation, and the evaluation of gate performance metrics, invariants, and observables. Optional classical noise can also be incorporated to assess gate performance and robustness under realistic operating conditions. 

Within this framework, the numerical implementation is optimized to mitigate the exponential scaling of the multi-qubit Hilbert space.
The computational cost, and in some cases the feasibility, of a simulation is determined by the choice of local Hilbert-space representation: the framework supports either the complete physical basis ($8^N$) or a reduced flip-flop subspace with lower dimensionality ($4^N$).

The complete $8^N$-dimensional representation retains all physical states of the system. To mitigate the exponential scaling bottleneck in multi-qubit simulations, we introduce a subspace approximation that restricts each electron-nucleus donor system to the states relevant for flip-flop dynamics. The reduced basis retains only configurations in which the electron and nuclear spins are anti-parallel, for both orbital configurations:
\begin{equation}
\{\ket{g\downarrow\Uparrow}, \ket{e\downarrow\Uparrow}, \ket{g\uparrow\Downarrow}, \ket{e\uparrow\Downarrow}\}.
\end{equation}
The excluded states correspond to parallel electron-nuclear spin configurations. Under the strong static magnetic field conditions established for  flip-flop operation, these configurations are energetically detuned and weakly coupled to the driven dynamics, allowing them to be safely neglected for coherent evolution. All results presented in this work are obtained using this reduced flip-flop subspace representation.

The time evolution is computed on a uniform time grid using piecewise-constant propagation, where the propagator over each time interval is evaluated through matrix exponentiation (MATLAB \texttt{expm}(\,)),
\begin{equation}
\hat{U}(t + dt) = \exp\left( - 2 \pi i \hat{H}(t) dt \right) \hat{U}(t),
\end{equation}
ensuring unitary evolution up to numerical precision.

\printbibliography[
heading=bibintoc,
title={References}
]

@article{Pedersen-2007,
title = {Fidelity of quantum operations},
author = {Line Hjortshøj Pedersen and Niels Martin Møller and Klaus Mølmer},
journal = {Physics Letters A},
volume = {367},
number = {1--2},
pages = {47--51},
year = {2007},
doi = {10.1016/j.physleta.2007.02.069},
}

@article{Makhlin-2002,
title = {Nonlocal Properties of Two-Qubit Gates and Mixed States, and the Optimization of Quantum Computations},
author = {Makhlin, Yuriy},
journal = {Quantum Information Processing},
volume = {1},
number = {4},
pages = {243--252},
year = {2002},
doi = {10.1023/A:1022144002391},
}

@article{Muhonen-2014,
  author    = {Muhonen, Juha T. and Dehollain, Juan P. and Laucht, Arne and Hudson, Fay E. and Kalra, Rachpon and Sekiguchi, Takeharu and Itoh, Kohei M. and Jamieson, David N. and McCallum, Jeffrey C. and Dzurak, Andrew S. and Morello, Andrea},
  title     = {Storing quantum information for 30 seconds in a nanoelectronic device},
  journal   = {Nature Nanotechnology},
  volume    = {9},
  pages     = {986--991},
  year      = {2014},
  doi       = {10.1038/nnano.2014.211}
}

@article{Petta-2005,
  author    = {Petta, J. R. and Johnson, A. C. and Taylor, J. M. and Laird, E. A. and Yacoby, A. and Lukin, M. D. and Marcus, C. M. and Hanson, M. P. and Gossard, A. C.},
  title     = {Coherent Manipulation of Coupled Electron Spins in Semiconductor Quantum Dots},
  journal   = {Science},
  volume    = {309},
  number    = {5744},
  pages     = {2180--2184},
  year      = {2005},
  doi       = {10.1126/science.1116955},
}

@article{Tyryshkin-2012,
  author    = {Tyryshkin, Alexei M. and Tojo, Shinichi and Morton, John J. L. and Riemann, Helge and Abrosimov, Nikolai V. and Becker, Peter and Pohl, Hans-Joachim and Schenkel, Thomas and Thewalt, Michael L. W. and Itoh, Kohei M. and Lyon, S. A.},
  title     = {Electron spin coherence exceeding seconds in high-purity silicon},
  journal   = {Nature Materials},
  volume    = {11},
  pages     = {143--147},
  year      = {2012},
  doi       = {10.1038/nmat3182},
  doi       = {10.1038/nmat3182},
}

@article{Pla-2012,
  author    = {Pla, J. J. and Tan, K. Y. and Dehollain, J. P. and Lim, W. H. and Morton, J. J. L. and Jamieson, D. N. and McCallum, J. C. and Dzurak, A. S. and Morello, A.},
  title     = {A single-atom electron spin qubit in silicon},
  journal   = {Nature},
  volume    = {489},
  number    = {7417},
  pages     = {541--545},
  year      = {2012},
  doi       = {10.1038/nature11449},
}

@article{Kane-1998,
  author    = {Kane, B. E.},
  title     = {A silicon-based nuclear spin quantum computer},
  journal   = {Nature},
  volume    = {393},
  pages     = {133--137},
  year      = {1998},
  doi       = {10.1038/30156},
}

@article{Pla-2013,
  author    = {Pla, Jarryd J. and Tan, Kuan Y. and Dehollain, Juan P. and Lim, Wee H. and Morton, John J. L. and Zwanenburg, Floris A. and Jamieson, David N. and Dzurak, Andrew S. and Morello, Andrea},
  title     = {High-fidelity readout and control of a nuclear spin qubit in silicon},
  journal   = {Nature},
  volume    = {496},
  number    = {7445},
  pages     = {334--338},
  year      = {2013},
  doi       = {10.1038/nature12011},
}

@article{Laucht-2015,
author = {Arne Laucht and Juha T. Muhonen and Fahd A. Mohiyaddin and Rachpon Kalra and Juan P. Dehollain and Solomon Freer and Fay E. Hudson and Menno Veldhorst and Rajib Rahman and Gerhard Klimeck and Kohei M. Itoh and David N. Jamieson and Jeffrey C. McCallum and Andrew S. Dzurak and Andrea Morello},
journal = {Science Advances},
pages = {e1500022},
title = {Electrically controlling single-spin qubits in a continuous microwave field},
volume = {1},
year = {2015},
doi = {10.1126/sciadv.1500022},
}

@article{Vandersypen-2017,
author = {L. M. K. Vandersypen and H. Bluhm and J. S. Clarke and A. S. Dzurak and R. Ishihara and A. Morello and D. J. Reilly and L. R. Schreiber and M. Veldhorst},
journal = {Npj Quantum Information},
pages = {34},
title = {Interfacing spin qubits in quantum dots and donors—hot, dense, and coherent},
volume = {3},
year = {2017},
doi = {10.1038/s41534-017-0038-y},
}

@article{Ferraro-2022,
author = {E. Ferraro and D. Rei and M. Paris and M. {De Michielis}},
journal = {EPJ Quantum Technology},
pages = {},
title = {Universal set of quantum gates for the flip-flop qubit with 1/f noise},
volume = {9},
number = {2},
year = {2022},
doi = {10.1140/epjqt/s40507-022-00120-7},
}

@article{Tosi-2017,
author = {Guilherme Tosi and Fahd A. Mohiyaddin and Vivien Schmitt and Stefanie Tenberg and Rajib Rahman and Gerhard Klimeck and Andrea Morello},
journal = {Nature Communications},
pages = {450},
title = {Silicon quantum processor with robust long-distance qubit couplings},
volume = {8},
year = {2017},
doi = {10.1038/s41467-017-00378-x},
}

@article{Calderon-2022,
author = {F. A. Calderon-Vargas and Edwin Barnes and Sophia E. Economou},
journal = {Phys. Rev. B},
pages = {165302},
title = {Fast high-fidelity single-qubit gates for flip-flop qubits in silicon},
volume = {106},
year = {2022},
doi = {10.1103/PhysRevB.106.165302},
}

@article{Simon-2020,
  title = {Fast noise-resistant control of donor nuclear spin qubits in silicon},
  author = {Simon, James and Calderon-Vargas, F. A. and Barnes, Edwin and Economou, Sophia E.},
  journal = {Phys. Rev. B},
  volume = {101},
  issue = {20},
  pages = {205307},
  numpages = {12},
  year = {2020},
  month = {May},
  publisher = {American Physical Society},
  doi = {10.1103/PhysRevB.101.205307},
}

@article{Tosi-2018,
author = {Guilherme Tosi and Fahd A. Mohiyaddin and Stefanie Tenberg and Arne Laucht and Andrea Morello},
journal = {Physical Review B},
pages = {075313},
title = {Robust electric dipole transition at microwave frequencies for nuclear spin qubits in silicon},
volume = {98},
year = {2018},
doi = {10.1103/PhysRevB.98.075313},
}

@article{Schuch-2003,
author = {Norbert Schuch and Jens Siewert},
journal = {Physical Review A},
pages = {032301},
title = {Natural two-qubit gate for quantum computation using the XY interaction},
volume = {67},
year = {2003},
doi = {10.1103/PhysRevA.67.032301},
}

@article{Steger-2012,
  author    = {Steger, M. and Saeedi, K. and Thewalt, M. L. W. and Morton, J. J. L. and Riemann, H. and Abrosimov, N. V. and Becker, P. and Pohl, H.-J.},
  title     = {Quantum Information Storage for over 180 s Using Donor Spins in a $^{28}$Si "Semiconductor Vacuum"},
  journal   = {Science},
  volume    = {336},
  number    = {6086},
  pages     = {1280--1283},
  year      = {2012},
  doi       = {10.1126/science.1217635},
}

@article{Morello-2020,
author = {A. Morello and J. J. Pla and P. Bertet and D. N. Jamieson},
journal = {Advanced Quantum Technologies},
pages = {2000005},
title = {Donor spins in silicon for quantum technologies},
volume = {3},
year = {2020},
doi = {10.1002/qute.202000005},
}

@article{McCallum-2021,
author = {J. C. McCallum and B. C. Johnson and T. Botzem},
journal = {Applied Physics Reviews},
pages = {031314},
title = {Donor-based qubits for quantum computing in silicon},
volume = {8},
year = {2021},
doi = {10.1063/5.0060957},
}

@article{Burkard-RevModPhys2023,
author = {Guido Burkard and Thaddeus D. Ladd and Andrew Pan and John M. Nichol and Jason R. Petta},
journal = {Rev. Mod. Phys.},
pages = {025003},
title = {Semiconductor Spin Qubits},
volume = {95},
year = {2023},
doi = {10.1103/RevModPhys.95.025003},
}

@article{Tenberg2019SpinRelaxationMOS,
  author    = {Tenberg, Stefanie B. and Asaad, Serwan and Mądzik, Mateusz T. and Johnson, Mark A. I. and Joecker, Benjamin and Laucht, Arne and Hudson, Fay E. and Itoh, Kohei M. and Jakob, A. Malwin and Morello, Andrea and Dzurak, Andrew S.},
  title     = {Electron spin relaxation of single phosphorus donors in metal-oxide-semiconductor nanoscale devices},
  journal   = {Physical Review B},
  volume    = {99},
  pages     = {205306},
  year      = {2019},
  doi       = {10.1103/PhysRevB.99.205306},
}

@article{Savytskyy2023Science,
author = {Rostyslav Savytskyy  and Tim Botzem  and Irene Fernandez de Fuentes  and Benjamin Joecker  and Jarryd J. Pla  and Fay E. Hudson  and Kohei M. Itoh  and Alexander M. Jakob  and Brett C. Johnson  and David N. Jamieson  and Andrew S. Dzurak  and Andrea Morello },
title = {An electrically driven single-atom flip-flop qubit},
journal = {Science Advances},
volume = {9},
number = {6},
pages = {eadd9408},
year = {2023},
doi = {10.1126/sciadv.add9408},
}

@article{DeMichielis-JPhysD-2023,
author = {M. {De Michielis} and E. Ferraro and E. Prati and L. Hutin and B. Bertrand and E. Charbon and D. J. Ibberson and M. F. Gonzalez-Zalba},
title = {Silicon Spin Qubits from Laboratory to Industry},
journal = {Journal of Physics D: Applied Physics} ,
pages = {363001},
number = {},
volume = {56},
year = {2023},
doi = {10.1088/1361-6463/acd8c7},
}

@article{DeMichielis-AQT-2024,
author = {M. {De Michielis} and D. Rei and E. Ferraro},
journal = {Advanced Quantum Technologies},
title = {Parallel Gate Fidelity of Flip-Flop Qubits in Small 1D- and 2D-Arrays in a Noisy Environment},
volume = {7},
number = {},
year = {2024},
pages = {2300455},
doi = {10.1002/qute.202300455},
}

@article{DeMichielis_LA_SA_STA-AQT-2025,
author = {M. {De Michielis} and E. Ferraro},
journal = {Advanced Quantum Technologies},
title = {Impact of Parallel Gating on Gate Fidelities in Linear, Square, and Star Arrays of Noisy Flip-Flop Qubits},
volume = {8},
number = {1},
year = {2025},
pages = {2400341},
doi = {10.1002/qute.202400341},
}

@software{FlipFlopQSim-2026,
title = {FlipFlopQSim},
author = {Lorenzo D'Onofrio and Elena Ferraro and Marco De Michielis},
year = {2026},
version = {1.0.0},
publisher = {Zenodo},
doi = {10.5281/zenodo.21669875},
}

\end{document}